\documentclass[aps.prb,amsmath,amssymb,twocolumn,floatfix]{revtex4-2}
\usepackage{graphicx}
\usepackage{dcolumn}
\usepackage{bm}
\usepackage{xcolor}

\begin{document}

\title{Atypical vortex lattice and the magnetic penetration depth in superconducting Sr$_2$RuO$_4$ deduced by $\mu$SR}
\author{M.~Yakovlev,$^1$ Z.~Kartsonas,$^2$ and J. E.~Sonier$^1$}

\affiliation{$^1$Department of Physics, Simon Fraser University, Burnaby, British Columbia V5A 1S6, Canada \\
$^2$Department of Physics and Astronomy, University of British Columbia, Vancouver, British Columbia V6T 1Z1, Canada}  

\date{\today}
\begin{abstract}
The muon spin rotation ($\mu$SR) technique has been applied to determine the behavior of the in-plane magnetic 
penetration depth ($\lambda_{ab}$) in the vortex state of the unconventional superconductor Sr$_2$RuO$_4$ as a means of 
gaining insight into its still unknown superconducting order parameter. A recent $\mu$SR study of Sr$_2$RuO$_4$ reported a $T$-linear temperature 
dependence for $\lambda_{ab}$ at low temperatures that was not identified in an earlier $\mu$SR study. 
Here we show that there is no significant difference between the data in the early and recent $\mu$SR studies and both are compatible with
the limiting low-temperature $\lambda_{ab} \! \propto \! T^2$ dependence expected from measurements of  
$\Delta \lambda_{ab}(T) \! = \! \lambda_{ab}(T) \! - \! \lambda_{ab}(0)$ in the Meissner state by other techniques.
However, we argue that at this time there is no valid theoretical model for reliablly determining the absolute value of $\lambda_{ab}$ in Sr$_2$RuO$_4$ 
from $\mu$SR measurements. Instead, we identify the formation of an unusual square vortex lattice   
that introduces a new constraint on candidate superconducting order parameters for Sr$_2$RuO$_4$.
\end{abstract}
\maketitle
\section{Introduction}
It has now been three decades since Sr$_2$RuO$_4$ was found to exhibit superconductivity \cite{Maeno:94}, yet a vital clue to
identifying the mechanism of Cooper pairing, namely the symmetry of the superconducting order parameter, has not been pinned down.
For the longest time Sr$_2$RuO$_4$ was believed to be a spin-triplet chiral $p$-wave (odd-parity) superconductor, primarily based on
the observation of an unchanging spin susceptibillity across $T_c$ \cite{Ishida:98,Duffy:00} 
and experimental evidence for broken time-reversal symmetry (TRS) \cite{Luke:98,Xia:06}.
In recent years revised nuclear magnetic resonance (NMR) Knight shift \cite{Pustgow:19, Ishida:20, Chronister:21} and 
polarized neutron scattering \cite{Petsch:20} 
measurements have reclassified the order parameter of the bulk superconducting state of Sr$_2$RuO$_4$ as ``even-parity spin-singlet''.
Moreover, broken TRS superconductivity in Sr$_2$RuO$_4$ has become controversial, with some experiments unable to confirm 
violation of this symmetry \cite{Kashiwaya:19} and proposals suggesting the cause may be of extrinsic origin \cite{Willa:21, Andersen:24}.
   
Measurements of the temperature dependence of the magnetic penetration depth ($\lambda$) are one way to probe the low-energy
quasiparticle excitations, gaining insight into the superconducting gap structure that is a manifestation of the pairing symmetry.
Experiments that measure the change in the in-plane magnetic penetration depth, 
$\Delta \lambda_{ab}(T) \! = \! \lambda_{ab}(T) \! - \! \lambda_{ab}(0)$, 
in the Meissner state predominantly observe $\Delta \lambda_{ab}(T) \! \propto \! T^2$ 
for Sr$_2$RuO$_4$ \cite{Bonalde:00, Ormeno:06, Mueller:24, Landaeta:24}. 
This has been taken as evidence for the existence of 
line nodes in the superconducting gap, under the assumption that the measurements at very low temperatures are influenced by a nonlocal 
electromagnetic response \cite{Kosztin:97,Roising:24}.
 
Recently, Khasanov {\it et al.} provided evidence for nodes in the superconducting gap based on transverse-field (TF) $\mu$SR measurements
of the temperature and magnetic field dependences of the in-plane magnetic penetration depth $\lambda_{ab}(T, B)$ in the vortex state 
of Sr$_2$RuO$_4$ single crystals ($T_c \! = \! 1.4$~K) \cite{Khasanov:23}. In particular, $\lambda_{ab}(T, B)$ was reported to 
exhibit a $T$-linear temperature dependence at low temperatures, in contrast to findings from an earlier TF-$\mu$SR study of 
Sr$_2$RuO$_4$ single crystals having a similar $T_c$ value of 1.45~K at zero magnetic field \cite{Luke:00}. 
In addition, $\lambda_{ab}(0, B)$ was found to exhibit a field 
dependence consistent with gap nodes. Here we show that the different dependence of $\lambda_{ab}$ on temperature 
reported in the early and recent TF-$\mu$SR studies of Sr$_2$RuO$_4$ has nothing to do with differences in sample quality, but 
rather how $\lambda_{ab}$ is derived from the TF-$\mu$SR asymmetry spectrum. 
We show that the limiting low-$T$ dependence of $\lambda_{ab}$ on temperature for the two samples are
actually similar and not inconsistent with the $T^2$ behavior of $\Delta \lambda_{ab}(T)$ determined by other experimental techniques that probe 
the Meissner state. In addition, we show that a vortex lattice with a
highly unusual spatial variation of magnetic field is necessary to describe the TF-$\mu$SR signal in Sr$_2$RuO$_4$, placing
a major constraint on its still undetermined unconventional superconducting order parameter. 

\section{Data analysis methods}
In what follows we compare four different kinds of data analysis methods for determining $\lambda_{ab}(T, B)$ in the vortex state of 
Sr$_2$RuO$_4$ from TF-$\mu$SR measurements:\\
\subsection{Second moment of the internal magnetic field distribution}
In Ref.~\cite{Khasanov:23} $\lambda_{ab}$ was determined using the following equation derived from Ginzburg-Landau (GL)
theory for the square root of the second moment of the magnetic field distribution $n(B)$ of an ideal vortex lattice (VL)  \cite{Brandt:03}  
\begin{eqnarray}
\langle \Delta B \rangle_{\rm VL}^{1/2}[\mu {\rm s}^{-1}] & = & A (1 - b)[1+ 1.21 \nonumber \\
& & \times (1 - \sqrt{b})^3] \lambda_{ab}^{-2}[\mu {\rm m}^{-2}] \, ,
\label{eqn:SecondMom}
\end{eqnarray}
where $A \! \simeq \! 4.83$ and 5.07 for an hexagonal and square VL, respectively \cite{Brandt:03,Khasanov:08}, and $b \! = \! B/B^{\parallel}_{c2}$ is the reduced 
field where $B^{\parallel}_{c2}$ is the upper critical magnetic field for a field applied parallel to the $c$ axis. For this orientation of applied field a square VL 
rotated 45$^\circ$ relative to the Ru-O-Ru bond directions in the $a$-$b$ plane was observed by small-angle neutron scattering in a $T_c \! = \! 1.28(6)$~K 
sample \cite{Riseman:98,Riseman:00}.

To obtain $\langle \Delta B \rangle_{\rm VL}^{1/2}$ the TF-$\mu$SR asymmetry spectra in Ref.~\cite{Khasanov:23} were fit to 
the following sum of Gaussian-damped cosine functions
\begin{eqnarray}
A(t) & = & \sum_{i=1}^{n} a_i e^{-\sigma_i^2 t^2} \cos(\gamma_\mu B_i t + \phi) \nonumber \\ 
       &    & + a_{\rm bg} e^{-\sigma_{\rm bg}^2 t^2} \cos(\gamma_\mu B_{\rm bg} t + \phi) \, ,
\label{eqn:Asymmetry}
\end{eqnarray} 
where $n \! = \! 2$, $\gamma_\mu \! = \! 851.615$~MHz/T is the muon gyromagnetic ratio, 
and the last term in Eq.~(\ref{eqn:Asymmetry}) accounts for the contribution to the TF-$\mu$SR time spectrum from muons stopping outside the sample.
The fitted values of the amplitude $a_i$, depolarization rate $\sigma_i$ and internal magnetic field $B_i$ for the 
$n \! = \! 2$ components of the sample term were subsequently used to calculate $\langle \Delta B \rangle_{\rm VL}^{1/2}$.
Here we point out that Eq.~(\ref{eqn:SecondMom}) is an approximation that is strictly valid for $\kappa \! = \! \lambda_{ab}/\xi_{ab} \! \geq \! 5$ and 
$\! 0.25/\kappa^{1.3} \! \lesssim \! b \! \ll \! 1$, where $\xi_{ab}$ is the in-plane GL coherence 
length \cite{Brandt:03}. For the Sr$_2$RuO$_4$ single crystals studied in Ref.~\cite{Khasanov:23}, 
$B^{\parallel}_{c2} \! \sim \! 75$~mT, which corresponds to $\xi_{ab} \! \sim \! 663$~\AA~ calculated 
from the GL relation $B^{\parallel}_{c2} \! =\! \Phi_0/2 \pi \xi_{ab}^2$.
This implies that Eq.~(\ref{eqn:SecondMom}) is valid provided $\lambda_{ab} \! > \! 3315$~\AA. Yet this is far greater than
the zero-temperature values of $1240$~\AA~$\lesssim \! \lambda_{ab}(0, b) \! \lesssim 1740$~\AA~ for $0 \! \leq \! b \! \leq \! 0.6$
reported in Ref.~\cite{Khasanov:23}. In addtion to this inconsistency, Eq.~(\ref{eqn:SecondMom}) is derived from a spatial field profile 
$B({\bf r})$ that does not account for the finite size of the vortex cores and assumes the individual vortices have a circular cross section,
despite the assumption of a square VL.
\begin{figure}
\centering
\includegraphics[width=8cm]{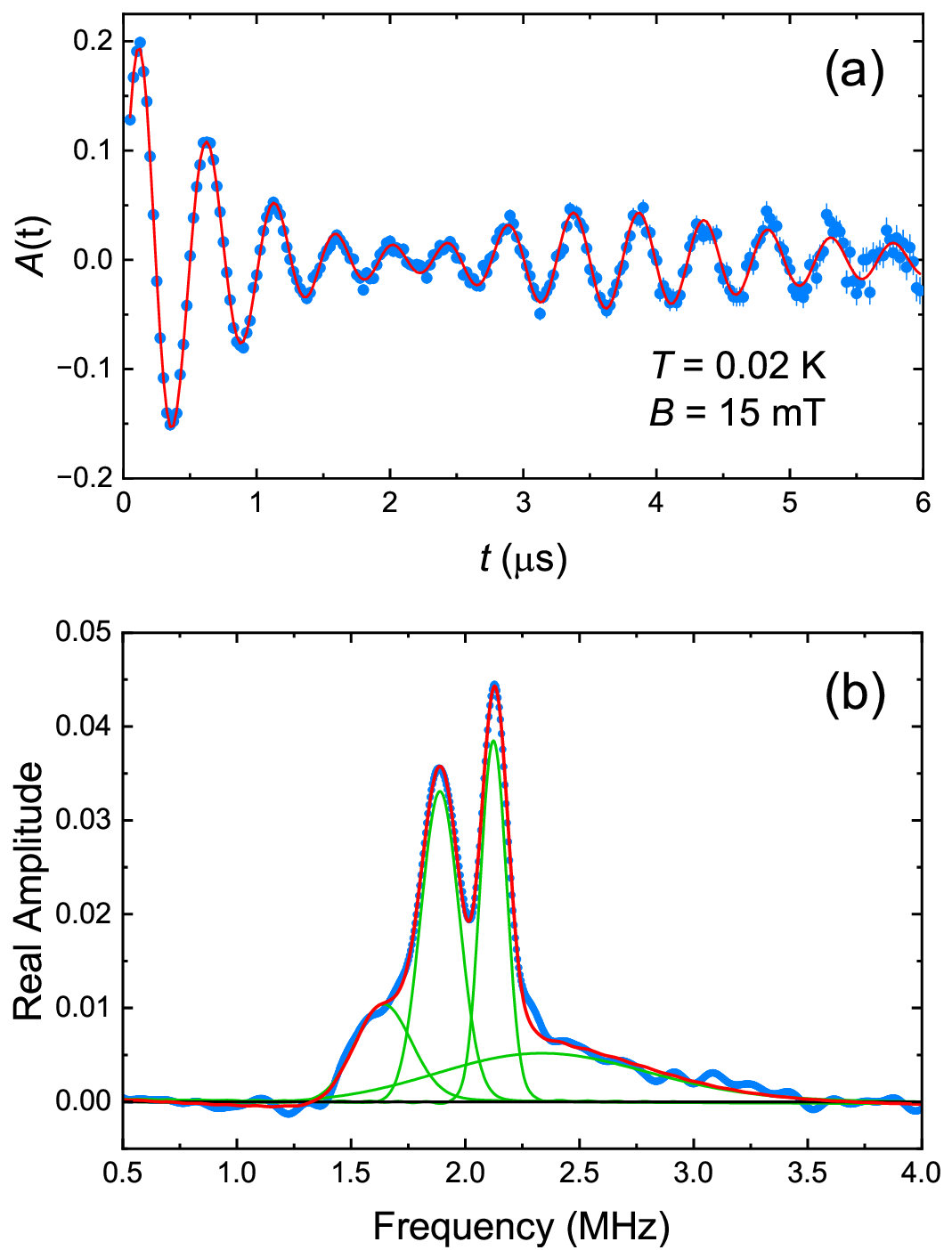}
\caption{(a) TF-$\mu$SR asymmetry spectrum recorded after field cooling the Sr$_2$RuO$_4$ single crystals of Ref.~\cite{Luke:00} to 
$T \! = \! 0.02$~K in a magnetic field of $B \! = \! 15$~mT applied parallel to the $c$ axis. The solid red curve through the blue data points
is a fit to Eq.~(\ref{eqn:Asymmetry}) for $n \! = \! 3$.
(b) Fourier transforms of the TF-$\mu$SR signal and of the fit in the time domain. The horizontal axis refers to the Larmor precession 
frequency of the muon spin in the local magnetic field $B$ at the muon site, given by $f_\mu \! = \! (\gamma_\mu/2 \pi)B$.
The peak near 2.12~MHz is the background contribution.
Also shown are Fourier transforms of the four individual Gaussian-damped cosine functions of the complete fit function.}
\label{fig1}
\end{figure}  

Before discussing the analysis method applied in the earlier TF-$\mu$SR study of \cite{Luke:00}, 
we present results obtained from applying the same analysis method of Ref.~\cite{Khasanov:23} to the earlier data.
Figure~\ref{fig1}(a) shows a representative TF-$\mu$SR asymmetry spectrum from the earlier study and a fit to Eq.~(\ref{eqn:Asymmetry}).
A noteable difference in the analysis of the TF-$\mu$SR spectra of the sample studied in Ref.~\cite{Luke:00}
is that $n \! = \! 3$ sample components are required to achieve a good fit. The need for an additional Gaussian-damped cosine function is apparent in the Fourier transform of the TF-$\mu$SR spectrum, which exhibits a low-frequency shoulder [see Fig.~\ref{fig1}(b)] that is not visually apparent in the
sample recently studied by Khasanov {\it et al.} (see Fig.~1(d) in Ref.~\cite{Khasanov:23}).
The fit parameters from this additional component ($a_3$, $\sigma_3$ and $B_3$) are included in our calculation of 
$\langle \Delta B \rangle_{\rm VL}^{1/2}$ and as in Ref.~\cite{Khasanov:23} this is converted to $\lambda_{ab}$ via Eq.~(\ref{eqn:SecondMom})
with $A \! = \! 5.07$. We note that the Fourier transform of the TF-$\mu$SR asymmetry spectrum provides a visual approximation of $n(B)$
associated with the VL and the distribution of field sensed by muons that missed the sample.
\begin{figure}
\centering
\includegraphics[width=8cm]{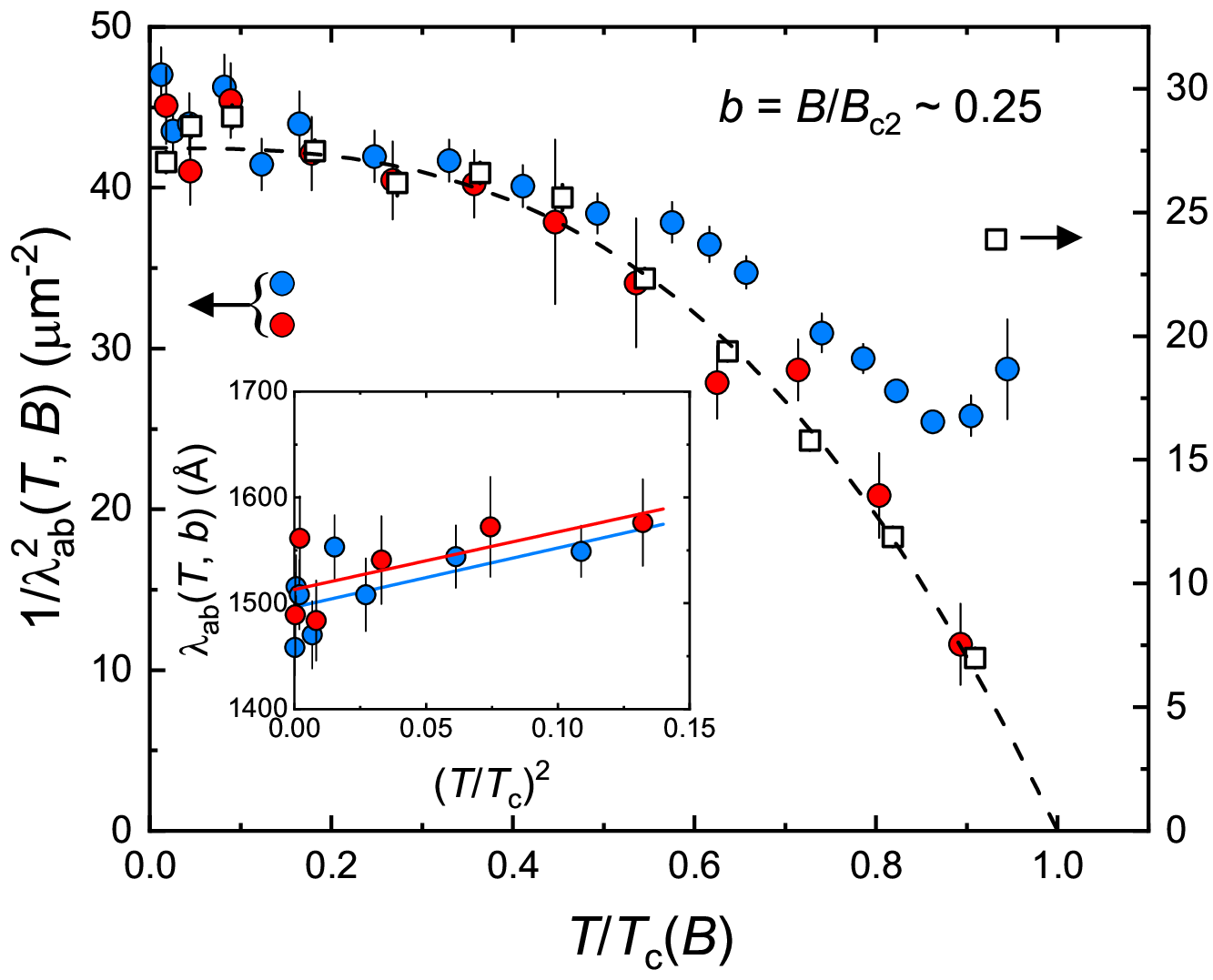}
\caption{The dependence of $1/\lambda_{ab}^2$ on the reduced temperature $T/T_c(B)$ for $b \! = \! B/B^{\parallel}_{c2} \! \sim \! 0.25$.
The solid data points are $1/\lambda_{ab}^2$ calculated from Eq.~(\ref{eqn:SecondMom}) for the TF-$\mu$SR measurements 
at $B \! = \! 20$~mT in Ref~\cite{Khasanov:23}, where $T_c(B) \! = \! 1.22$~K (blue circles), and at $B \! = \! 15$~mT in Ref~\cite{Luke:00}, 
where $T_c(B) \! = \! 1.13$~K (red circles). 
The vertical scale on the right corresponds to the open square data points and the dashed fit curve 
[{\it i.e.}, $1/\lambda_{ab}^2(T) \! \propto \! (1 - (T/T_c)^{2.78})$] from Ref.~\cite{Luke:00}. 
The inset shows $\lambda_{ab}$ vs. $(T/T_c)^2$ for $T \! \leq \! 0.36 T_c$, where the solid lines are linear fits to the two data sets.} 
\label{fig2}
\end{figure}

Figure~\ref{fig2} shows how $1/\lambda_{ab}^2$ vs. $T/T_c$ from this kind of data analysis compares to the results
of Ref.~\cite{Khasanov:23} for a similar reduced field $b$.
Despite the additional component required to fit the TF-$\mu$SR asymmetry spectra for the earlier sample \cite{Luke:00}, 
there is remarkable agreement with the behavior and values of $1/\lambda_{ab}^2(T, b)$ 
reported in Ref.~\cite{Khasanov:23} below $T \! \sim \! 0.4 T_c$.
Yet while the temperature dependence of $1/\lambda_{ab}^2$ determined from $\langle \Delta B \rangle_{\rm VL}^{1/2}$
for the earlier sample closely follows that initially reported in Ref.~\cite{Luke:00} over the entire range,
the data from Ref.~\cite{Khasanov:23} departs from this behavior above  $T \! \sim \! 0.4 T_c$.
This is likley a consequence of the more symmetric magnetic field distribution observed in the 
vortex state of the sample studied in Ref.~\cite{Khasanov:23}. The frequencies of the multiple Gaussian-damped
cosine components become less separated as the asymmetry of the internal magnetic field distribution descreases with
increasing temperature, so that indentifiable fit parameters are more difficult to achieve. On the other hand, the low-field
shoulder of the field distribution in the older sample is clearly observed in the Fourier transform of the TF-$\mu$SR
signal up to $T \! = \! 0.7T_c$ (see Fig.~2 in Ref.~\cite{Luke:00}, so the sample-component frequencies remain distinct. 
The absence of the third lower-frequency sample component in the single crystals investigated in Ref.~\cite{Khasanov:23}
indicates a significant difference in the VL magnetic field profile $B({\bf r})$ and suggests
an hexagonal rather than square arrangement of the vortices in this sample.
Regardless, the calculated values of $1/\lambda_{ab}^2(0, b)$ from Eq.~(\ref{eqn:SecondMom}) for $A \!= \! 4.83$ and $A \!= \! 5.07$ differ by 
only about $5$~\%. 

The $T$-linear temperature dependence of $\lambda_{ab}$ reported in Ref.~\cite{Khasanov:23} was 
inferred from a linear fit of $\lambda_{ab}(T,B)$ extending up to $T \! \sim \! 0.57 T_c$, which is far above temperatures at which quasiparticle 
excitations are confined to nodal regions on the Fermi surface. 
The inset of Fig.~\ref{fig2} shows that the variation of $\lambda_{ab}$ with temperature derived 
from $\langle \Delta B \rangle_{\rm VL}^{1/2}$ for both samples is actually not incompatible with the $T^2$ temperature dependence of $\Delta \lambda_{ab}(T)$ observed in the Meissner state by other experimental techniques. At the very least, the scatter in the data prevents reliably
distinguishing between a $T$-linear and $T^2$ temperature dependence. 

An alternative way to determine the absolute value and behavior of
$\lambda_{ab}(T, B)$ is to fit the TF-$\mu$SR spectra to a theoretical model for the field profile $B({\bf r})$ of the VL \cite{Sonier:00}.
This approach includes multiplying the sample contribution to $A(t)$ by Gaussian depolarization functions to account for
additional broadening of the internal magnetic field distribution by the nuclear dipole moments and VL disorder, and like the
second moment analysis, adding a residual background term (Gaussian-damped cosine function) to account for muons that missed
the sample, as follows
\begin{eqnarray}
A(t) & = & a_{\rm s}e^{-(\sigma_{\rm_n}^2 + \sigma_{\rm dis}^2) t^2} \sum_{ {\bf r}}\cos \left[ \gamma_\mu B({\bf r})  t + \phi \right]  \nonumber \\
       &    & + a_{\rm bg} e^{-\sigma_{\rm bg}^2 t^2} \cos(\gamma_\mu B_{\rm bg} t + \phi) \, .
\label{eqn:ModelAsymmetry}
\end{eqnarray} 
The parameter $\sigma_{\rm n}$ is the depolarization rate due to the nuclear dipoles in the sample, which is independent of temperature and determined
from fits of TF-$\mu$SR asymmetry spectra recorded above $T_c$. The depolarization rate $\sigma_{\rm dis}$ accounts for further broadening of 
the internal magnetic field distribution by frozen VL disorder, which is dependent on temperature and assumed to be proportional 
to $1/\lambda_{ab}^2(T)$ \cite{Sonier:07}.
The sum in the first term is over real-space positions {\bf r} in an ideal periodic VL. Below, we compare attempts to 
use Eq.~(\ref{eqn:ModelAsymmetry}) to analyze the Sr$_2$RuO$_4$ TF-$\mu$SR asymmetry spectra of Ref.~\cite{Luke:00} 
for three different models of $B({\bf r})$.
\begin{figure}
\centering  
\includegraphics[width=7.7cm]{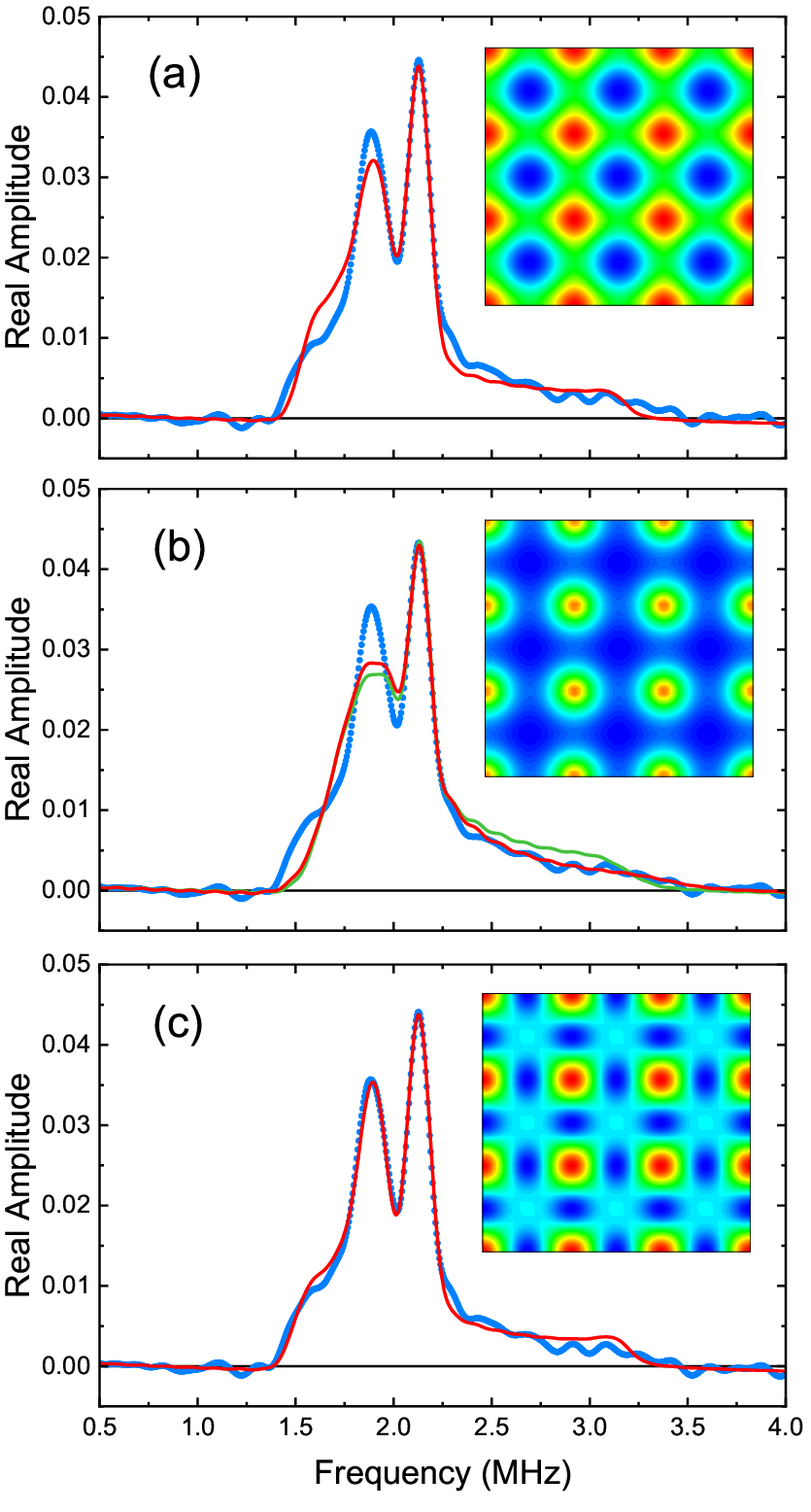}
\caption{Fourier transforms of the TF-$\mu$SR asymmetry spectrum of Sr$_2$RuO$_4$ from Ref.~\cite{Luke:00} for
$T \! = \! 0.02$~K and $B \! = \! 15$~mT (blue data points) and unconstrained fits of the TF-$\mu$SR spectrum 
(red curves) to (a) the modified 
nonlocal London model, yielding $\lambda_{ab} \! = \! 567(8)$~\AA, $\kappa \! = \! 0.74(1)$~\AA, $n_{xxyy} \! = \! 0$,
and $d \! = \! 20(1)$,
(b) the iterative GL model,  yielding $\lambda_{ab} \! = \! 1832(41)$~\AA and $\kappa \! = \! 6.4(7)$~\AA, and  
(c) the two-component $E_u$ state model, yielding $\lambda_{ab} \! = \! 1790(9)$~\AA,
$\xi_{ab} \! = \! 2.65(7)$~\AA~ and Fermi surface anisotropy $\nu \! = \! 0.31(1)$. 
Also shown in (b) is a fit to the iterative GL model with $\kappa$ fixed to 2.06 (green curve). The contour plots in each panel
depict $B({\bf r})$ for the VL projected onto the $a$-$b$ plane generated by the unconstrained fits.
The intervortex spacing in each contour plot is 373~nm and
 the red and dark blue colors correspond to the maximum and minimum values of $B({\bf r})$, respectively.} 
\label{fig3}
\end{figure}       
\subsection{Nonlocal London model} 
Since Sr$_2$RuO$_4$ is a low-$\kappa$ type-II superconductor, non-local effects are presumably important for describing the VL induced by 
the applied field. Kogan {\it et al.} have made nonlocal corrections to the London equations, which 
couple the VL to the crystal anisotropy of the Fermi velocity and accounts for the hexagonal-to-square VL transition that occurs  
in V$_3$Si and borocarbide superconductors at high field \cite{Kogan:97a,Kogan:97b}. 
Multiplying the nonlocal London solution by an anisotropic cutoff factor derived within GL theory to account 
for finite size of the vortex cores results in the following equation for $B({\bf r})$ that has been successfully used to determine $\lambda_{ab}$
and $\xi_{ab}$ from TF-$\mu$SR measurements of V$_3$Si through the hexagonal-to-square VL transition \cite{Sonier:04}
\begin{equation}
B({\bf r}) = B  \sum_{ {\bf G}}
\frac{(1-b^4)e^{-i {\bf G} \cdot {\bf r}} \, \, u \, K_1(u)}{\lambda_{ab}^2 G^2 + 
\lambda_{ab}^4(n_{xxyy} \, G^4 + d \, G_x^2 G_y^2)} \, .
\label{eq:NLm}
\end{equation}
Here $B$ is the mean internal magnetic field, 
{\bf G} are the reciprocal lattice vectors of the VL, $u^2 \! = \! 2 \xi_{ab}^2 G^2 (1 + b^4)[1-2b(1 - b)^2]$, $K_1(u)$ is a modified Bessel function, 
and $n_{xxyy}$ and $d$ are dimensionless parameters arising from nonlocal corrections.
As defined earlier, $b \! = \! B/B^{\parallel}_{c2}$. The  $a$-$b$ plane GL coherence length is a fit parameter for this model, but the
upper critical field is not, since they are related by the GL relation  $B^{\parallel}_{c2} \! = \! \Phi_0/2 \pi \xi_{ab}^2$.

We have attempted to fit the TF-$\mu$SR spectra of the Sr$_2$RuO$_4$ sample 
from Ref.~\cite{Luke:00} to an asymmetry function that assumes this modified nonlocal London model of $B({\bf r})$ 
for a square VL. 
As shown in Fig.~\ref{fig3}(a), this model poorly describes the low-field shoulder and left peak of the
Fourier transform of the TF-$\mu$SR signal, which correspond to the minimum and saddle point magnetic fields of $B({\bf r})$, respectively.
The failure of the nonlocal London model to adequately describe the VL of Sr$_2$RuO$_4$ is not 
unexpected, since this model is strictly valid for high-$\kappa$ superconductors.\\

\subsection{Iterative GL model} 
Brandt developed an iteration method for solving the GL equations to accurately determine $B({\bf r})$
for $b \! \geq \! 10^{-3}$, $\kappa \! \geq \! 1/\sqrt{2}$, and aribtrary VL symmetry \cite{Brandt:97}.
This model has previously been used to determine $\lambda(T, b)$ and $\xi(T, b)$ from TF-$\mu$SR measurements of the hexagonal VL in
the low-$\kappa$ ($\kappa \! \sim \! 1.3$) type-II superconductor V \cite{Laulajainen:06}.
For a detailed description of the iteration method to compute $B({\bf r})$, see Refs.~\cite{Brandt:97,Laulajainen:06}.
The values of $b$ and $\kappa$ for which this iterative GL model is valid make it applicable to the TF-$\mu$SR studies of Sr$_2$RuO$_4$.
Nevertheless, good fits of the TF-$\mu$SR spectra of Ref.~\cite{Luke:00} could not be achieved with this model for $B({\bf r})$
and best fits yield an unrealistic value for $\kappa$ (see Table~\ref{Table1}). Fixing $\kappa$ to 2.06, which corresponds to the calculated ratio
$\lambda_{ab}/\xi_{ab}$, where $\lambda_{ab}$ is the $T \! = \! 0$ extrapolated value from the second moment analysis (see Fig.~\ref{fig2} inset)
and $\xi_{ab}$ is calculated from the GL equation for $B^{\parallel}_{c2}$, leads to a worse fit [see Fig.~\ref{fig3}(b)].
Like the modified nonlocal London model, the iterative GL model fails to describe the low-field shoulder and the left peak
of the Fourier transformed TF-$\mu$SR signal.\\

 \begin{table}
\caption{Fit parameters and reduced $\chi^2$ for the different fits of the TF-$\mu$SR asymmetry spectrum of Sr$_2$RuO$_4$ 
for $T \! = \! 0.02$~K and $b \! \sim \! 0.25$  from Ref.~\cite{Luke:00}. Fourier transforms of the TF-$\mu$SR signal and the fits 
are displayed in Fig.~\ref{fig3}.}
\begin{ruledtabular}
\begin{tabular}{lccc}
{\bf Model} & $\kappa$ & $\lambda_{ab}$ (\AA) & Reduced $\chi^2$ \\ \hline
Second moment & - & 1490(65) & 1.077 \\
Modified nonlocal London &  0.74(1) & 567(8) & 1.936 \\
Iterative GL &  6.4(7) & 1832(41)  & 2.576 \\
                   &  2.06 (fixed) & 1386(10) & 3.336 \\
Two-component $E_u$ state& 2.65(7) & 1790(9) & 1.281\\
\end{tabular}
\end{ruledtabular}
\label{Table1}
\end{table}

\subsection{Two-component $E_u$ state model} 
The temperature dependence of $\lambda_{ab}$ in Sr$_2$RuO$_4$ for $b \! \sim \! 0.25$ 
was determined in Ref.~\cite{Luke:00} from a global fit of the TF-$\mu$SR spectra to a model of $B({\bf r})$ 
derived from GL theory for a two-component complex order parameter belonging to the two-dimensional irreducible 
representation $E_u$ of the tetragonal point group $D_{4h}$ \cite{Agterberg:98}.
The breaking of TRS in the superconducting state requires a multiple-component order parameter, and this has long favored the
spin-triplet chiral $p$-wave pairing state ${\bf d}({\bf k}) \! = \! \Delta_0(k_x \! \pm \! i k_y)\hat{z}$ 
belonging to the $E_u$ representation. The two-component $E_u$ state model predicts 
the square VL observed in Sr$_2$RuO$_4$ by small-angle neutron scattering \cite{Riseman:98,Riseman:00} and is valid for low $\kappa$. 
The components of the analytical equation for $B({\bf r})$ in this two-component $E_u$ state model are given by 
numerous equations and hence we refer the reader to Ref.~\cite{Agterberg:98} for full details. The
model includes the GL-parameter $\kappa$ and a Fermi surface anisotropy parameter $\nu$, which were treated
as common temperature-independent fit parameters in the data analysis for Ref.~\cite{Luke:00}.
It is visually apparent in Fig.~\ref{fig3}(c) and clear from a comparison of the goodness-of-fit of the TF-$\mu$SR
asymmetry spectrum for $T \! = \! 0.02$~K summarized in Table~\ref{Table1} that this model closely 
resembles $B({\bf r})$ for the VL in Sr$_2$RuO$_4$.
As shown in Fig.~\ref{fig2}, the temperature dependence of $1/\lambda_{ab}^2$ originally determined from the global fit analysis 
assuming this model \cite{Luke:00} qualitatively agrees with that determined from $\langle \Delta B \rangle_{\rm VL}^{1/2}$, 
although the absolute value of $\lambda_{ab}$ is about 30~\% larger. 

The contour plot in Fig.~\ref{fig3}(c) shows that $B({\bf r})$ for the VL generated by a fit to the two-component $E_u$ state model
corresponds to an unusual situation where the minimum field is midway between nearest-neighbor vortices, rather than at the center 
of the square VL unit cell. While it now seems the superconducting order parameter of Sr$_2$RuO$_4$ does not belong to the $E_u$ representation,
as the revised NMR Knight shift measurements indicate even-parity superconductivity, the unconventional
location of the minimum field of the square VL is telling. In particular, this unique characteristic of the VL 
must be a manifestation of the true superconducting order parameter.

\section{Conclusions}
To summarize, we have shown by a common analysis method that there is no significant difference between the temperature dependence of 
the in-plane magnetic penetration depth $\lambda_{ab}$ for Sr$_2$RuO$_4$ determined in early and recent TF-$\mu$SR studies,
but that there is too much scatter in both data sets to support the conclusion 
in Ref.~\cite{Khasanov:23} of a limiting low-temperature $\lambda_{ab} \! \propto \! T$ dependence. Instead, 
the data from both the early and recent studies reasonably follow a limiting $\lambda_{ab} \! \propto \! T^2$ dependence,
in agreement with measurements of $\Delta \lambda_{ab}(T)$ in the Meissner state by other experimental techniques.
In the current work we have also demonstrated that there is at the present time no valid model for determining the absolute value of $\lambda_{ab}$ 
in Sr$_2$RuO$_4$ from TF-$\mu$SR measurements.
 
Lastly, we have identified a unique feature of the spatial field profile of the VL in Sr$_2$RuO$_4$, which
introduces a new constraint for candidate superconducting order parameters.
Here we mention a nodal $d_{x^2 - y^2}$ order parameter, which remains a leading candidate for Sr$_2$RuO$_4$ and
for which a $\lambda_{ab} \! \propto \! T^2$ dependence due to nonlocal electrodynamics is expected in a low-$\kappa$ superconductor \cite{Kosztin:97,Roising:24}. 
For a square VL rotated in the $a$-$b$ plane 45$^{\circ}$ with respect to the $a$ axis, which is the case for Sr$_2$RuO$_4$ \cite{Riseman:98,Riseman:00},
at high $b$ the fourfold symmetry around the vortex cores due to a $d_{x^2 - y^2}$ order parameter
enhances $B({\bf r})$ at the center of the unit cell in the next-nearest neighbor direction and suppresses
 $B({\bf r})$ at the saddle point between nearest-neighbor vortices \cite{Ichioka:99}. This situation is similar to
the contour plot in Fig.~\ref{fig3}(c).  

\begin{acknowledgments}
J.E.S. acknowledges support from the Natural Sciences and Engineering Research Council (NSERC) of 
Canada (PIN: 146772). We thank David Broun, Graeme Luke and Daniel Agterberg for informative discussions.
\end{acknowledgments}

\end{document}